\documentclass[twocolumn,showpacs,preprintnumbers,amsmath,amssymb,superscriptaddress]{revtex4}
\usepackage{graphicx}
\usepackage{dcolumn}
\usepackage{bm}
\usepackage{longtable}
\usepackage[countmax]{subfloat}
\usepackage{epstopdf}
\usepackage{color}

\newcommand{\h}{h}

\newcommand{\meanv}[1]{\left\langle#1\right\rangle}

\newcommand{\E}{\mathbb{E}}

\providecommand{\be}{\begin{equation}}
  \providecommand{\ee}{\end{equation}}
\providecommand{\bea}{\begin{eqnarray}}
  \providecommand{\eea}{\end{eqnarray}}
\providecommand{\beas}{\begin{eqnarray*}}
  \providecommand{\eeas}{\end{eqnarray*}}

\providecommand{\beni}{\begin{equation*}}
  \providecommand{\eeni}{\end{equation*}}

\providecommand{\bw}{\begin{widetext}}
  \providecommand{\ew}{\end{widetext}}

\def\a{\alpha}
\def\s{\sigma}

\def\b{\beta}

\def\d{\delta}

\usepackage{epsfig}
\usepackage{amssymb}
\usepackage{amsmath}
\usepackage{amsthm}
\usepackage{latexsym}
\usepackage{graphicx}
\def\be{\begin{equation}}
\def\ee{\end{equation}}
\def\bc{\begin{center}}
\def\ec{\end{center}}

\def\be{\begin{equation}}
\def\ee{\end{equation}}
\def\bea{\begin{eqnarray}}
\def\eea{\end{eqnarray}}

\begin{document}

\preprint{APS/123-QED}

\title{Phase transitions in Restricted Boltzmann Machines with generic priors}

\author{Adriano Barra}
\affiliation{Dipartimento di Matematica e Fisica Ennio De Giorgi, Universit\`a del Salento, Lecce, Italy}
\author{Giuseppe Genovese}
\affiliation{Institut f\"ur Mathematik, Universit\"at Z\"urich, Z\"urich, Switzerland.}
\author{Peter Sollich}
\affiliation{Department of Mathematics, King College London, London, U.K.}
\author{Daniele Tantari}
\affiliation{Centro Ennio De Giorgi, Scuola Normale Superiore, Pisa, Italy.}

\date{\today}

\begin{abstract}
We study Generalised Restricted Boltzmann Machines with generic priors for units and weights, interpolating between Boolean and Gaussian variables. We present a complete analysis of the replica symmetric phase diagram of these systems, which can be regarded as Generalised Hopfield models. We underline the role of the retrieval phase for both inference and learning processes and we show that retrieval is robust for a large class of weight and unit priors, beyond the standard Hopfield scenario. Furthermore we show how the paramagnetic phase boundary is directly related to the optimal size of the training set necessary for good generalisation in a teacher-student scenario of unsupervised learning. \end{abstract}


\maketitle

In recent years supervised machine learning with neural networks has found renewed interest from the practical success of so-called \textit{deep networks} in solving several difficult problems, ranging from image classification to speech recognition  and video segmentation \cite{deep}.
Despite this remarkable progress, unsupervised learning with neural networks, in which the structure of data is learned without a priori knowledge of a specific task,  still lacks a solid theoretical scaffold. Such learning of hidden features of complex data in high dimensional spaces by fitting a generative probabilistic model is used for denoising, completion and data generation, but also as a dimensionality reduction pre-training step in supervised methods \cite{ben2,hintrev}. 
 
In this framework, given a set of parameters $\boldsymbol{\xi}$, a probability density $P(\boldsymbol{\s},\boldsymbol{h}|\boldsymbol{\xi})$ over the joint space of the visible units $\boldsymbol{\s}$ (representing the data) and a set of hidden units  $\boldsymbol{h}$ is introduced. \textit{Learning} is the process of determining an optimal set of parameters by fitting the marginal distribution $P(\boldsymbol{\s}|\boldsymbol{\xi})$ to the data, e.g.\ through likelihood maximization \cite{hintmomentum,engel,benbook,gabrie1,parigi}. 
In the \textit{inference} process, once optimal parameters are learned, hidden units become selectively activated by data features through $P(\boldsymbol{h}|\boldsymbol{\s},\boldsymbol{\xi})$. How many data are necessary to learn a set of statistically relevant features (learning reliability), which features of the data hidden units respond to  and how strongly (inference reliability), clearly depends on the dataset but also on the choice of the generative model used  \cite{ben2,hintrev}.

Hereafter we focus on  \textit{generalized Restricted Boltzmann Machines} ({\em RBM})  \cite{smole, hinton2, freund, hinton1}, defined by $N$ visible units $\{\s_i\}_{i=1}^{N}$ and $P$ hidden units $\{h_\mu\}_{\mu=1}^{P}$ whose joint probability density is parameterized by an  $N\times P$ matrix $\boldsymbol{\xi}$ as
 \be\label{eq:joint}
 P(\boldsymbol{\s},\boldsymbol{h}|\boldsymbol{\xi})= \frac{P_\s(\boldsymbol{\s})P_h(\boldsymbol{\h})e^{\sum_{i=1}^{N}\sum_{\mu=1}^{P} \xi^{\mu}_i \s_ih_\mu}}{Z(\boldsymbol{\xi})},
 \ee
where $P_\s$ and $P_h$ are generic priors and the partition function
$Z(\boldsymbol{\xi})$ is a normalisation factor.  We assume priors factorize over components so that a
RBM can be represented as an undirected \textit{bipartite} graph with two interacting layers of units and no connections within the same layer.  This property makes inference straightforward, since the conditional distribution over the hidden layer factorizes. Furthermore the marginal distribution over the visible layer can be exactly computed as 
\be\label{eq:marg}
P(\boldsymbol{\s}|\boldsymbol{\xi})=Z^{-1}(\boldsymbol{\xi})P_\s(\boldsymbol{\s})\exp\left(\sum_{\mu=1}^{P}u\left(\sum_{i=1}^{N}\xi^\mu_i\s_i\right)\right),
\ee
with $u(x)=\log\E_h e^{xh}$ the cumulant generating function of the hidden unit prior.  This defines a so-called \textit{Generalized Hopfield Model} ({\em GHM}). This class includes for  $u(x)=x^2/2$ (i.e.\ a Gaussian hidden prior) the standard Hopfield model, introduced in the context of pattern recognition \cite{hopfield1}, whose $P$ patterns are the columns of the weight matrix $\boldsymbol{\xi}$. 


  RBMs have been widely used to model probability distributions over binary data, for which they are universal approximators \cite{ben1}. Important progress has been made in the last few years in using RBMs with different priors in both visible and hidden layers, going beyond the Boolean case to better capture features in the data \cite{hinton3,courv1,courv2}.

Our aim in this paper is to study the phase diagram of the bipartite model (\ref{eq:joint}), by exploiting the equivalence with GHMs  (\ref{eq:marg}). In particular the existence and robustness of a retrieval phase in the GHM is intimately related to the significance and representational power of the RBM hidden units as features explaining the data distribution. Moreover the analysis of the transition of the GHM from a paramagnetic to a spin glass phase helps to determine the size of the training set necessary in the RBM for a good estimate of the data distribution.

Driven by these motivations, we analyze the phase diagram of a random RBM when the priors of the units (visible and hidden) and weights are of the general form
\be\label{eq:prior}
P_\Omega(\eta)\propto \sum_{\epsilon=\pm 1} \exp(-(\eta-\sqrt{1-\Omega}\epsilon)^2/2\Omega)
\ee
that interpolates between a binary $\pm1$ $(\Omega \to 0)$ and a Gaussian $(\Omega \to 1)$ distribution, drawing both the visible and hidden units
independently from this prior, i.e.\ $P_\s(\boldsymbol{\s})=\prod_i P_{\Omega_\s}(\s_i)$,  $P_h(\boldsymbol{h})=\prod_\mu P_{\Omega_h}(h^\mu)$. We study the standard fully connected case of weak patterns, rescaling $\xi^{\mu}_i\to \sqrt{\b/N}\xi^\mu_i$,
and set $P_\xi(\boldsymbol{\xi})=\prod_{i,\mu} P_{\Omega_\xi}(\xi^\mu_i)$. The relative intensity $\b>0$ of the patterns  w.r.t.\ the units plays the role of the \textit{inverse temperature} in a
statistical mechanics language.  We introduce $\d=\sqrt{1-\Omega_\xi}$ to specify the position of the two peaks in the pattern prior and study the asymptotic phase diagram of (\ref{eq:joint}) in the limit of large $N$, varying  $\b$, $\boldsymbol{\Omega}=(\Omega_\sigma,\Omega_h,\Omega_\xi$ [or $\d$]$)$ and the relative size of the two layers, $\a=P/N$.

\begin{figure}
\includegraphics[scale=.6]{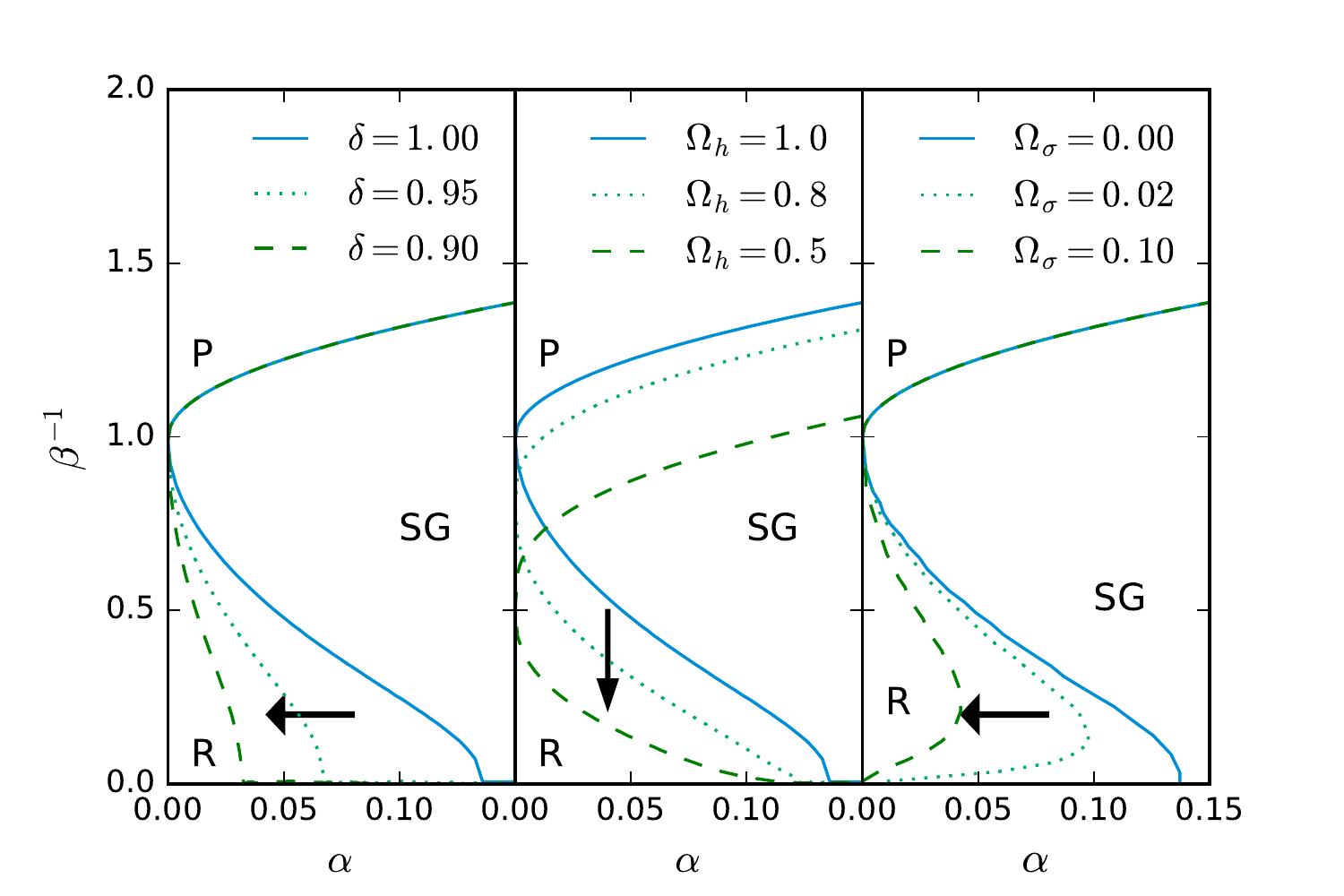}
\caption{($\a,\b^{-1}$) RS phase diagram of a random RBM for  varying pattern, hidden and visible unit priors (resp.\ $\d$, $\Omega_h$, $\Omega_\s$). Left: Boolean visible units and $\Omega_h=1$; the retrieval region approaches the line $\a=0$, $\b^{-1}\in[0,\Omega_h]$ as $\d\to 0$. Middle: Boolean visible units and $\d=1$; the retrieval region approaches the line $\b^{-1}=0$, $\a\in[0,\a_c(\d)]$ as $\Omega_h\to 0$. Right: $\d=1$, $\Omega_h=1$ and soft visible units regularized with a spherical constraint. The retrieval region shrinks to low load as $\Omega_\s\to 1$. In the three plots the blue line is for the standard Hopfield model ($\Omega_\s=0$, $\Omega_\tau=\d=1$), which turns out to have the largest retrieval region.}\label{fig1}
\end{figure}

The Boltzmann probability distributions $(\ref{eq:joint})$ and ($\ref{eq:marg}$) can be studied by using as order parameters the overlaps with the patterns, i.e. the magnetizations
\be
 m^\mu
 =\frac 1 {N} \sum_{i=1}^{N}\xi^\mu_i \s_i,
\ee
and the two replica overlaps 
\be
q^{ab}=\frac 1 {N} \sum_{i=1}^{N}\s^a_i\s^b_i, \quad  r^{ab}=\frac{1}{P} \sum_{\mu=1}^{P} h^a_\mu h^b_\mu.
\ee
The superscripts $a$, $b$ indicate (for $a\neq b$) two independent configurations drawn from $(\ref{eq:joint})$; the case $a=b$ defines the self-overlaps,  denoted as $Q\equiv q^{aa}$ and $R\equiv r^{aa}$.
In a \textit{replica symmetric} treatment \cite{amit,engel,CKS}, all overlaps are self-averaging as $N\to\infty$, with values $m^\mu$, $q$, $r$, $Q$ and $R$ satisfying (see \cite{lungo} for a detailed derivation)
\be
m^\mu = \meanv{\xi^\mu \meanv{\s}_{\s|z,\xi}}_{z,\xi}
\label{eqs:1}
\ee
\begin{eqnarray}
q    = \meanv{ \meanv{\s}^2_{\s|z,\xi}}_{z,\xi} &\quad& r =   \meanv{ \meanv{h}^2_{h |\eta}}_{\eta,\xi}\label{eqs:3} \\
Q         = \meanv{ \meanv{\s^2}_{\s|z,\xi}}_{z,\xi} &\quad& R    =   \meanv{ \meanv{h^2}_{h |\eta}}_{\eta,\xi}. \label{eqs:5}
\end{eqnarray}
Here $z,\eta\sim \mathcal{N}(0,1)$ and the entries of $\xi$ are sampled from $P_\xi$; the distributions of $\s$ and $h$ are proportional to
\begin{eqnarray}
&& P_{\Omega_\s}(\s) e^{\b\Omega_h (\boldsymbol{m}\cdot\xi) \s +\sqrt{ \b\a r} z\s+\frac{\b\a}{2} (R-r)\s^2},\label{eq:effdistr1}\\
&& P_{\Omega_h}(h) e^{ \sqrt{\b q}h \eta +\frac {\b} {2}(Q-q) h^2 }.\label{eq:effdistr2}
\end{eqnarray}
For weak patterns ($\b\ll1$) one has a paramagnetic phase where the distributions (\ref{eq:effdistr1}), (\ref{eq:effdistr2}) have zero effective fields $m$, $q$ and $r$. On increasing $\beta$ a transition to a spin glass phase takes place, with frozen disordered states that have $m=0$ but non-zero $q$ and $r$. Assuming the transition to be continuous, linearizing ($\ref{eqs:3}$) around zero gives the transition criterion
\be\label{eq:tc}
1= \b^2 \a  \meanv{\s^2}_0^2 \meanv{h^2}_0^2=\b^2 \a Q^2R^2\,.
\ee
Here $\meanv{}_0$ denotes the expectation value w.r.t.~(\ref{eq:effdistr1}) and (\ref{eq:effdistr2}) for $q=r=0$; from ($\ref{eqs:5}$), $Q$ and $R$ solve
\begin{eqnarray}
Q         &=&  \meanv{\s^2}_0 =  (1-\b\a\Omega_\s^2R)/(1-\b\a\Omega_\s R)^2\,\label{eqQ}\\
R      &=&   \meanv{h^2}_0 =  (1-\b\Omega_h^2Q)/(1-\b\Omega_h Q)^2.\nonumber
\end{eqnarray}
We remark that this result is independent of the \emph{pattern} prior, while it does depend on
the unit priors. Combining $(\ref{eq:tc})$ and ($\ref{eqQ}$), one gets the paramagnetic-spin glass
transition line  $\b_{SG}(\a)$ (Fig.~$\ref{fig1}$). For Boolean visible units ($\Omega_\s=0$) one has explicitly 
\be\label{eq:psg}
\b^{-1}_{SG}(\a)=\Omega_h+\frac{\sqrt{\a}}{2}+\frac 1 2[\a+4\Omega_h(1-\Omega_h)\sqrt{\a}]^{1/2}, 
\ee
The leading term for large $\a$ is $\sqrt{\a}$, independently of $\Omega_h$, while for $\a\to 0$ one gets $\b^{-1}_{SG}(0)=\Omega_h$. Eq.~(\ref{eq:psg}) generalizes existing results for the bipartite Sherrington-Kirkpatrick ($\Omega_h=0$) and the standard Hopfield ($\Omega_{h}=1)$ models where respectively $\b^{-1}_{SG}=\sqrt{\a}$ and $\b^{-1}_{SG}=1+\sqrt{\a}$ \cite{bipartito1,bipartito2,bipartito3,bipartito4,bipartito5}.


When the intensity of the patterns increases ($\b\sim 1$), a ferromagnetic transition to non-zero magnetization $m^\mu$ may occur, in which the visible layer is significantly correlated with a pattern (retrieval).  A non-zero magnetization is associated with a high-signal and a low-noise random field in the distribution (\ref{eq:effdistr1}) and 
this requires the size of the second layer (or ``pattern load") to be relatively small.
In the low-load regime, i.e.\ $\a=0$, 
linearising for small $m^\mu$ gives $
\boldsymbol{m}=\b\Omega_h\boldsymbol{m} +O(\boldsymbol{m}^3)$  and a bifurcation occurs at $\b=\Omega_h^{-1}$, independently of the other  priors. In the high-load regime, i.e.\ $\a>0$, the critical point for retrieval $\b^{-1}_R(\a)$ starts from $\Omega_h$ for small $\alpha$, decreases and eventually vanishes at the critical load $\a_c$, beyond which ferromagnetic states cannot exist. The size and the shape of the retrieval region are dependent on the priors.

 If the first layer has $\pm1$ units ($\Omega_\s=0$), taking the limit $\b\to\infty$ in (\ref{eqs:1}-\ref{eqs:5}) one finds a first order phase transition in the overlap $m^\mu$ at $\a_c(\d)$. The latter
decays roughly exponentially with $1-\d$, and it is maximal when the prior forces the patterns to be $\pm1$.  The whole retrieval line $\b_R(\a)$ can be found numerically after some manipulation of (\ref{eqs:1}-\ref{eqs:5}). In the $(\a,\b^{-1})$ plane this line connects the two limit points $(\a_c,0)$ and $(0,\Omega_h)$: the retrieval region disappears when either $\d\to 0$ or $\Omega_h\to 0$ while the other is kept fixed (Fig.~\ref{fig1} left and middle).


Models with real variables in both layers, even if regularized as in (\ref{eq:prior}), are usually ill-defined for $\b> 1$, due to the occurrence of negative eigenvalues in the interaction matrix. 
 The singularity can be removed by adding a stronger regularization on at least one layer.
For instance we can add the spherical constraint $\d(N-\sum_{i=1}^{N} \s_i^2)$ to the prior $P_\s(\boldsymbol{\s})$ (which then still depends on $\Omega_\s$). At $\Omega_\s=1$ the prior is uniformly distributed on the sphere, at $\Omega_\s=0$ the constraint is irrelevant.  Following the same replica calculations, the only difference is an extra Gaussian tail $e^{-\omega \s^2/2}$ in the effective $\s$-unit distribution, with $\omega$ a Lagrange multiplier chosen to fix the radius $Q = 1$ \cite{lungo}. The results for these models indicate that the retrieval region is robust also in the high-load regime, disappearing as $\Omega_\s\to1$ (Fig.\ \ref{fig1} right).  Summarizing, the existence of the retrieval region is robust beyond the standard Hopfield model, which however is optimal amongst priors of the form ($\ref{eq:prior}$).  

A RBM in a retrieval phase models relatively few, highly probable, clusters of configurations $\boldsymbol{\sigma}$ that can be described in terms of their closeness to the patterns $\boldsymbol{\xi}^\mu$. In a paramagnetic phase, on the other hand, there is just one unstructured cluster, while in the spin glass phase an exponentially large number of clusters arises that are uncorrelated with the patterns. Often it is desirable for unsupervised learning to provide an \emph{interpretable} representation, in terms of relatively few relevant features that capture the heterogeneity of the data but avoid overfitting. For RBMs this requires that the system is in a retrieval region after training: only then can the learned patterns be regarded as a set of interpretable features. This feature status can be seen from the inferred distribution of the hidden unit 
$h^\mu$ for pattern
$\boldsymbol{\xi}^\mu$, 
\be
P(h^\mu|\boldsymbol{\s}, \boldsymbol{\xi})\propto P_h( h^{\mu}) \exp(N m^\mu(\boldsymbol{\s}) h^{\mu})
\ee
where the magnetization acts as an external field: in the retrieval phase, one (learning by prototypes \cite{hopf2}) or a few (learning by features \cite{mona}) hidden units are strongly activated while in the paramagnetic and spin glass phase all the hidden units are weakly and incoherently activated. During learning with unnormalized patterns the \textit{effective inverse temperature} $\Vert \boldsymbol{\xi}^{\mu}\Vert^2/N$ can decrease until the RBM reaches the retrieval region \cite{mona}. Our results (Fig.~\ref{fig1}) show that the chance of this occurring can be enhanced by appropriate choice of structural parameters, maximizing the retrieval region using appropriate unit priors ($\Omega_\s\ll 1$, $\Omega_h \approx 1$), weight regularization ($\delta \approx 1$) and size of the layers ($\alpha \ll 1$). 

 \begin{figure}
\includegraphics[scale=0.62]{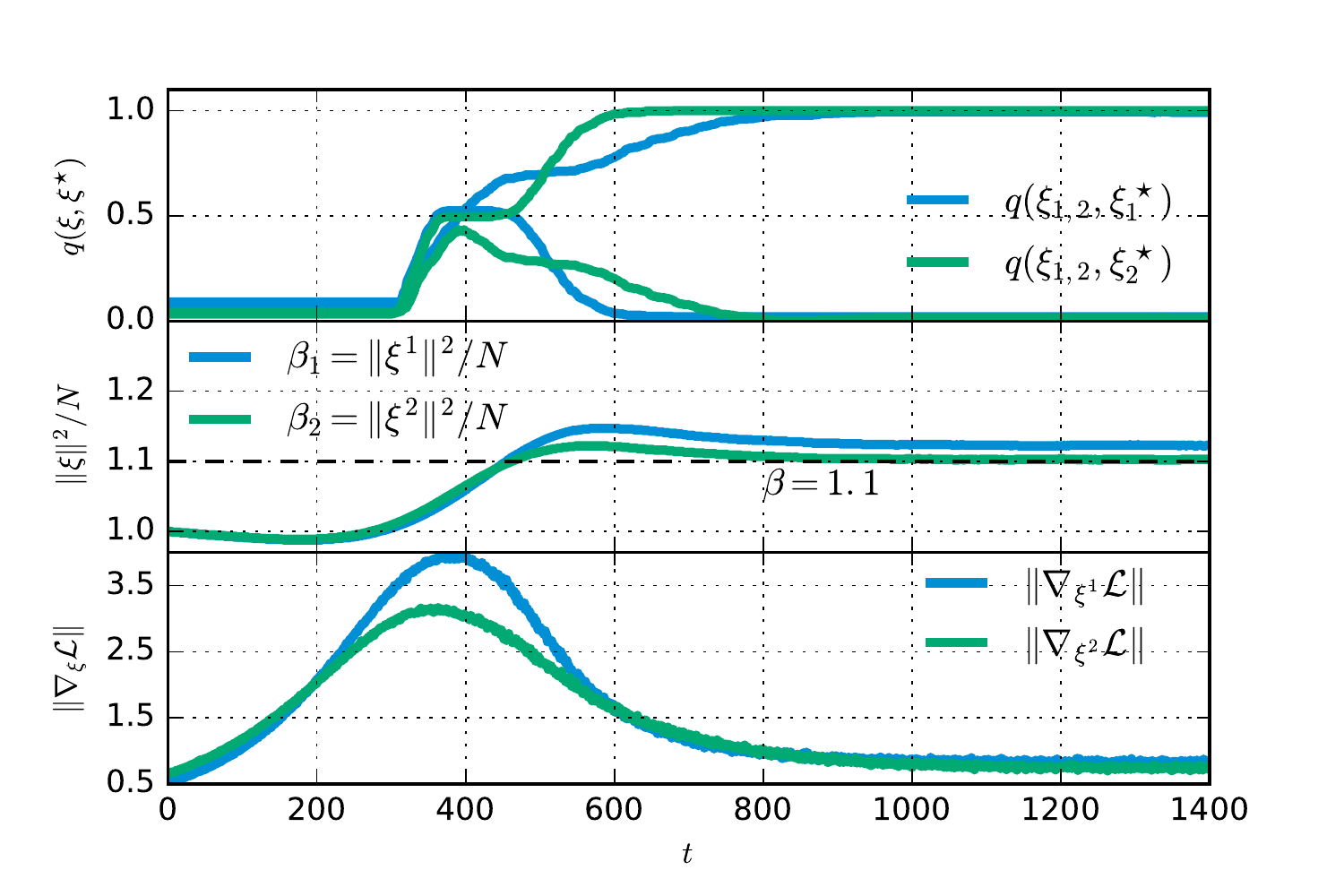}
\caption{Learning strong patterns ($\b>\Omega_h=1$) by training an RBM with $N=1000$ and $P=2$ from a training set of $M=100$ samples $\{\boldsymbol{\bar{\s}}^b\}_{b=1}^M$ supplied  by a teacher GHM.
Top: overlap between planted (teacher) and inferred (student) patterns. Middle: effective pattern temperatures tend to $\b$ during learning. Bottom: RBM training was performed by gradient descent on the likelihood $\mathcal{L}=-\ln P( \boldsymbol{\bar{\s}}|\boldsymbol{\xi})$, using Contrastive Divergence \cite{hintmomentum} to compute the gradient at each learning period $t$. 
\label{fig3}}
\end{figure}

\begin{figure}
\includegraphics[scale=.62]{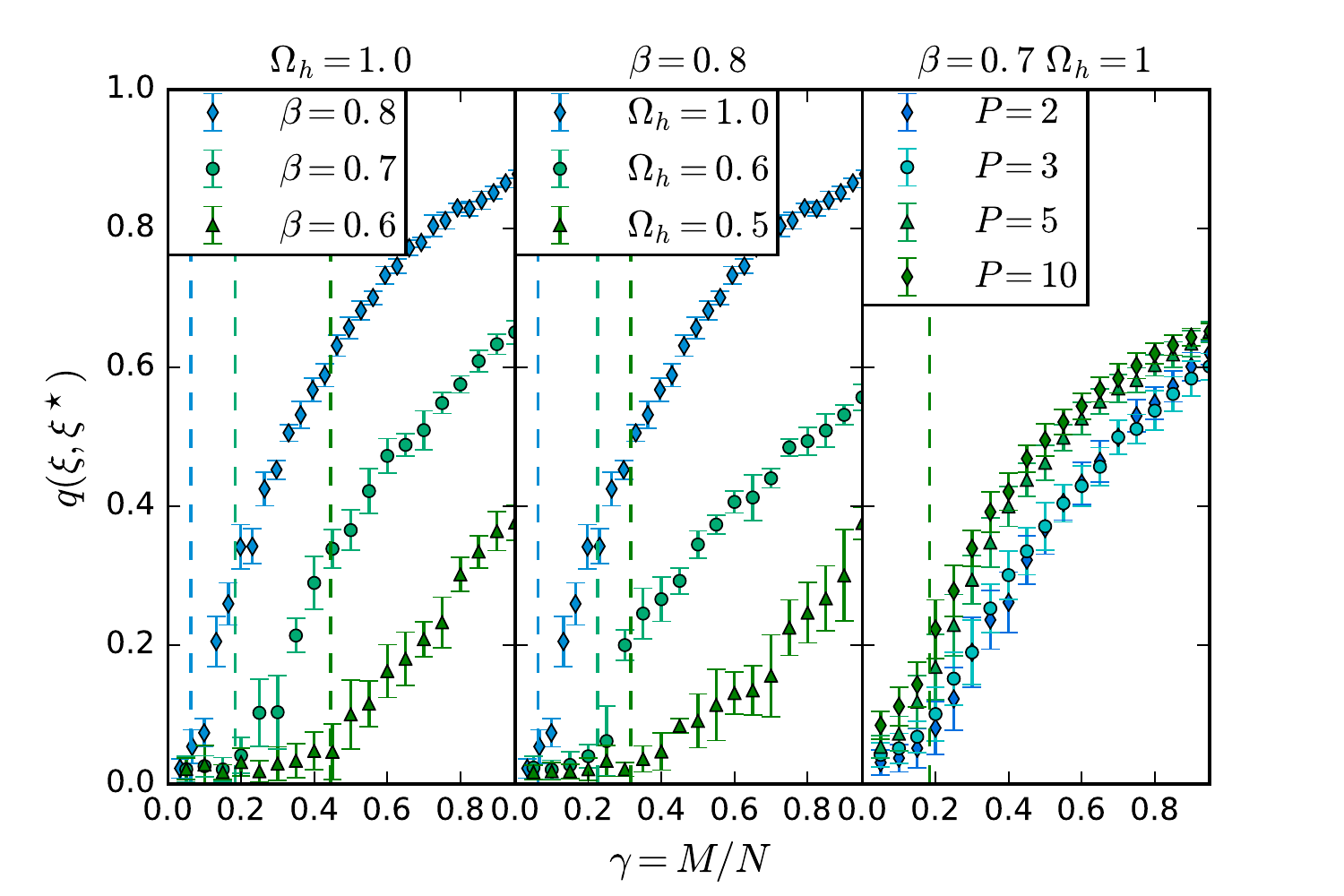}
\caption{ Overlap between planted ($\boldsymbol{\xi^\star}$) and inferred ($\boldsymbol{\xi}$) patterns in a teacher-student experiment for weak patterns ($\b<1$) as a function of the training set size $M=\gamma N$. 
Vertical lines indicate the theoretical detectability transitions. Left:  $P=1$, $\Omega_h=1$, $\gamma_c$ grows with decreasing $\b$. Middle: $P=1$, $\b=0.8$, $\gamma_c$ grows with decreasing $\Omega_h$  (low signalling hidden units). Right: $\b=0.7$, $\Omega_h=1$, $\gamma_c$ is effectively independent of $P$ as $P\ll N$.  Inferred $\boldsymbol{\xi}$ are obtained from a MC dynamics that samples from the posterior $(\ref{eq:posterior})$ with a system of $N=2000$ visible units.}\label{fig2}
\end{figure} 

Irrespective of interpretability, a more basic requirement for unsupervised learning 
is the ability to reproduce as accurately as possible a probability distribution over data, learning features from a relatively small training set: the more samples are needed, the less useful the model may be. Hence it is important to determine the minimum number of configurations in the training set required for good generalization.

A simple setting for this question is a teacher-student scenario, where data are produced by a teacher GHM with temperature $\b$ and \textit{planted} patterns $\boldsymbol{\xi}^\star$ unknown to the student. The student has to optimize the parameters of an RBM, learning as precisely as possible the original patterns from a training set of $M=\gamma N$ samples $\boldsymbol{\bar{\s}}=\{\boldsymbol{\bar{\s}}_b\}_{b=1}^M$ supplied by the teacher.

At low temperature, $\b^{-1}<\Omega_h$, and for $\alpha\ll 1$ (small second layer) the samples $\boldsymbol{\bar{\s}}$ are drawn from a GHM in the retrieval phase, thus different samples are strongly correlated with each other and with the original patterns $\boldsymbol{\xi^\star}$, making learning easy even with relatively few samples. In Fig.~\ref{fig3} the whole learning process is shown: a student RBM is trained 
on a training set of $M$ samples. \footnote{We use maximum likelihood training, which discards the pattern prior term, because the latter only has a weak effect in this regime.}
The student's patterns quickly converge toward the planted ones even using a small number of samples $M$: $\operatorname{sign}(\boldsymbol{\xi})$ becomes positively correlated with $\boldsymbol{\xi^\star}$ and the effective inverse temperature  $\Vert \boldsymbol{\xi}^\mu \Vert^2/N$ for each pattern approaches the teacher's $\b$. 

The problem is harder if planted patterns are weak ($\b^{-1}>\Omega_h$). It can be studied by introducing the posterior distribution over patterns, given the observed $\boldsymbol{\bar{\s}}$,
%
%
\begin{eqnarray}\label{eq:posterior}
P(\boldsymbol{\xi}|\boldsymbol{\bar{\s}})&=&\frac{P_\xi(\boldsymbol{\xi}) \prod_{b=1}^M P(\bar{\boldsymbol{\s}}^b|\boldsymbol{\xi})}{P(\bar{\boldsymbol{\s}})}  \propto W(\boldsymbol{\xi},\boldsymbol{\bar{\s}}):= \\
&=&Z^{-M}(\boldsymbol{\xi}) \prod_{\mu=1}^{P} P_\xi(\boldsymbol{\xi}^\mu) \exp \left( \sum_{b=1}^Mu\left(\sum_{i=1}^{N}\bar{\s}^b_i \xi^{\mu}_i\right) \right).\nonumber
\end{eqnarray}
The reliability of learning
can be measured by the average overlap
$\ q(\boldsymbol{\xi},\boldsymbol{\xi^\star})=\boldsymbol{\xi}\cdot\boldsymbol{\xi^\star}/(NP)$
between the
inferred and the true patterns:
\begin{eqnarray}
q &=& \int d\boldsymbol{\xi^\star}d\boldsymbol{\bar{\s}}\,d\boldsymbol{\xi}  \ q(\boldsymbol{\xi},\boldsymbol{\xi^\star}) \ P_\xi(\boldsymbol{\xi^\star}) P(\bar{\boldsymbol{\s}}|\boldsymbol{\xi^\star})P(\boldsymbol{\xi}|\boldsymbol{\bar{\s}})  \nonumber\\
&=& \left\langle \int d\boldsymbol{\xi}\,d\boldsymbol{\xi^\star}\, q(\boldsymbol{\xi},\boldsymbol{\xi^\star})  \ P(\boldsymbol{\xi}|\boldsymbol{\bar{\s}})P(\boldsymbol{\xi^\star}|\boldsymbol{\bar{\s}})\right\rangle_{P(\boldsymbol{\bar{\s}})}.
\end{eqnarray}
This is the overlap between two configurations drawn from the posterior $P(\cdot |\boldsymbol{\bar{\s}})$, averaged over $P(\boldsymbol{\bar{\s}})$. In this planted disorder setting,  $P(\boldsymbol{\bar{\s}})= P_\s(\boldsymbol{\bar{\s}})  \int d\boldsymbol{\xi}\, W(\boldsymbol{\xi}, \boldsymbol{\bar{\s}}) $ is defined in terms of the partition function itself \cite{parigi}. This differs from a \textit{quenched} model, with i.i.d.\ disorder $\boldsymbol{\bar{\s}}\sim P_\s(\boldsymbol{\bar{\s}})$. In the planted scenario, using the replica formalism,
\begin{eqnarray}
 q&=&   \int{d\boldsymbol{\xi}\,d\boldsymbol{\xi^\star}} \, q(\boldsymbol{\xi},\boldsymbol{\xi^\star})  \int{d\bar{\boldsymbol{\s}}} \frac{W(\boldsymbol{\xi},\boldsymbol{\bar{\s}}) W(\boldsymbol{\xi^\star},\boldsymbol{\bar{\s}})}{ \int d\boldsymbol{\hat{\xi}}\, W(\boldsymbol{\hat{\xi}}, \boldsymbol{\bar{\s}})}P_\s(\boldsymbol{\bar{\s}}) \nonumber \\
 &=&\lim_{n\to1} \int{d\boldsymbol{\xi}^1\cdots d\boldsymbol{\xi}^n} \ q(\boldsymbol{\xi}^1,\boldsymbol{\xi}^2) \meanv{\prod_{a=1}^n W(\boldsymbol{\xi}^a,\boldsymbol{\bar{\s}})}_{P_\s(\boldsymbol{\bar{\s}})}
 \end{eqnarray}
Here the limit $n\to1$ replaces the prescription $n\to 0$ for the quenched case.  If $q>0$, it is possible to learn patterns correlated with the original (planted) ones; for $q=0$ this is impossible. The two cases define two regions in the plane of $(\gamma,\b^{-1})$, divided by a line marking a detectability transition during learning.
As a general feature of this kind of systems, planted and quenched disorder are equivalent in the paramagnetic region $q=0$ up to the spin glass transition to $q>0$. The transition is the same, therefore, in the two cases \cite{howglassy,parigi} and can be obtained by studying the paramagnetic--spin glass transition of the \emph{quenched} model. 

In the simple case of a single Boolean pattern, $P=1$, $W(\boldsymbol{\xi},\boldsymbol{\bar{\s}})\propto\exp \left( \sum_{b=1}^Mu\left(\sum_{i=1}^{N}\bar{\s}^b_i \xi_i\right) \right)$  and our inference problem can be mapped to the study of a {\em dual} GHM, where the pattern $\boldsymbol{\xi}$ to be inferred plays the role of the {\em dual} spin configuration, and the sampled spin configurations $\boldsymbol{\bar{\s}}^b$ correspond to the {\em dual} patterns.
 At high temperature one then needs a minimum number of samples $M_c=\gamma_c (\b) N$ to avoid the paramagnetic region, i.e.\ to overcome the noise in the samples that are now very weakly correlated with the pattern \cite{mon1,engel}.  The function $\gamma_c(\b)$ is given by inverting the paramagnetic-spin glass transition line $\b_{SG}^{-1}(\gamma)$ of (\ref{eq:psg}) since in the dual model $\gamma$ plays exactly the role of the load $\a$ in the GHM.  The theoretical  $\gamma_c (\b)$ is in good agreement with the transition of the overlap between planted and inferred pattern found in a numerical learning  experiment, as shown in Fig.~$\ref{fig2}$.  As expected the critical training set size increases when $\b$ decreases and so weakens the signal from $\boldsymbol{\xi}^\star$. It also increases when $\Omega_h$ decreases as Fig.~\ref{fig1} (middle) shows, confirming again that hidden units with broader priors are able to supply higher signals. Inference was performed with a Monte Carlo (MC) dynamics over the posterior  ($\ref{eq:posterior}$), i.e.\ a GHM with $M$ patterns $\boldsymbol{\bar{\s}}^b$ sampled from a teacher GHM with planted $\boldsymbol{\xi}^\star$.  Note that the result is independent on the learning procedure used: the same transition can be obtained for example by training an RBM as in \cite{mona} or by a PCA analysis on the correlation matrix of the data \cite{mon1}.
 
 When $P>1$ the posterior $(\ref{eq:posterior})$ is the product of $P$ dual GHMs, one for each pattern to be inferred and all with the same dual patterns $\boldsymbol{\bar{\s}}^b$, coupled via a term $M\log Z(\boldsymbol{\xi})$ in the Hamiltonian. Without this term the detectability threshold would be the same as for a single GHM, i.e.\ the same as for $P=1$. For general $P\ll N$ it can  be shown that the term $Z^M(\boldsymbol{\xi})$ becomes a rigid constraint on to the subspace of mutually orthogonal $\boldsymbol{\xi}^\mu$: out of all the frozen states of the $P$ dual GHMs, which are otherwise independent, the constraint selects those in which the patterns are mutually orthogonal, preventing e.g.\ the inference of one of the $P$ planted patterns more than once. The analytical details supporting this argument will be given in a forthcoming work. However our numerical simulations (Fig.~\ref{fig2}) already point in this direction, showing that the detectability threshold is independent of $P$ when $P\ll N$. For larger $P$ this must break down; specifically for $P>M$ (where for binary $\boldsymbol{\sigma}$ and $\boldsymbol{\xi}$ the number of bits to learn exceeds the number of bits in the data), learning cannot be possible. Therefore it would be interesting to extend our analysis also to the case of $P$ growing with $N$, where we expect the detectability threshold to depend on the ratio $P/M$ as well as on $\a$.

Our study characterizes the structural conditions (low effective temperature, relative size of the layers, weight and unit priors) that allow RBMs to operate as interpretable feature extractors and to be learned accurately from data. It would be interesting to extend our work to architectures with more than one layer of hidden units, e.g.\ by considering more structured types of weight prior, as in \cite{mezard}. Moreover, one should consider also the possibility of tuning the weight sparsity \cite{soll1,prlgalluz,cool1,coolmedium,tso}. This has a crucial role in increasing the critical capacity for feature retrieval, but also in regulating the so-called compositional phase \cite{mona} where multiple features are extracted simultaneously.

\smallskip

\noindent {\bf Acknowledgments.} A.B. and D.T. acknowledges GNFM-INdAM for financial support via the grant Agliari2016 and Tantari2016. G.G. is supported by the NCCR SwissMAP.

\end{document}